\documentclass[12pt]{article}
\usepackage{latexsym}
\usepackage{graphicx}
\usepackage{multirow}
\usepackage{amsmath}
\usepackage{ulem}
\usepackage{setspace}
\usepackage{subfigure}
\begin{document}
\begin{center}\Large{Mass Generation via Gravitational confinement of fast Neutrinos}
\end{center}
\begin{center}
\renewcommand\thefootnote{\fnsymbol{footnote}}
Constantinos G. Vayenas$^{1,2,}$\footnote{E-mail: cgvayenas@upatras.gr.}, Dimitrios Grigoriou$^1$ and Dionysios Tsousis$^1$
\end{center}
\begin{center}
\textit{$^1$University of Patras, GR 26504 Patras, Greece\\$^2${Academy of Athens, GR-10679 Athens, Greece}}
\end{center}
\begin{abstract}
We analyze the dependence on neutrino energy of the gravitational attraction between ultrarelativistic neutrinos using Special Relativity and the equivalence principle of inertial and gravitational mass. It is found that when accounting for the special relativistic increase in particle mass with particle speed, then this gravitational force reaches the value of the Strong Force at a neutrino energy of 313 MeV, corresponding to the effective mass of a quark. This force can sustain the rotational motion of self-gravitating neutrino triads or electron/positron-neutrino pairs, and thus lead to the self driven generation of mass, i.e. of hadrons and bosons, the masses of which can be computed with an astonishing precision of 1\% without any adjustable parameters. The same final results are obtained via the Schwarzschild geodesics of General Relativity.\\
\vspace{0.5cm}
\end{abstract}

\section{Introduction}
The Standard Model of particle physics has long provided physicists with a basis for understanding the basic structure of all observable matter in the universe. Among the indivisible particles it
describes are quarks – which make up ‘hadrons’ including protons and neutrons; and leptons – which include electrons, and the far more elusive neutrinos. So far, the Standard Model has been an excellent basis for researchers to explain their observations \cite{Griffiths08,Tully11}, with experimental results rarely deviating far from theoretical calculations. Yet despite its success, physicists have known for some time that the model in its current form cannot be complete. It assumes that neutrinos are massless and also it appears to be incompatible with General Relativity.

The recent measurement of the pressure distribution in protons \cite{Burkert18} has emphasized the importance of gravitational interactions between quarks \cite{Vayenas12,Research,Vayenas13}. Such interactions inside protons have been reasonably considered until recently to be negligible in comparison with the Strong Force, yet both Special relativity (SR) \cite{Einstein1905,French68} coupled with the equivalence principle \cite{Vayenas12,Roll1964}, and also the use of the Schwarzschild geodesics of General Relativity (GR) \cite{Vayenas12,Vayenas13,Lomonosov2} have shown in recent years \cite{Vayenas12,Research} that the gravitational attraction between ultrarelativistic neutrinos with energies above 150 MeV is stronger than the electrostatic attraction of positron-electron pairs at the same distance and reaches the value of the Strong force for neutrino energies above 300 MeV. This very strong force can confine neutrinos to circular orbits with fm size radii and extremely high speeds corresponding to Lorentz factor $\gamma$ values of the order of 10$^{10}$ \cite{Vayenas12,Vayenas13,Lomonosov1,Vayenas20}. Such ultrarelativistic neutrinos have the energy, $\gamma m_oc^2$, of quarks, i.e. $\sim$ 300 MeV. Conversely, the rest mass, $m_o$, of quarks is in the rest mass range of neutrinos, i.e. up to $\sim$ 0.05 eV/c$^2$ \cite{Vayenas12,Vayenas13,Mohapatra07}.

Such simple considerations have led to the development of the Rotating Lepton Model (RLM) \cite{Vayenas12,Research,Vayenas13,Vayenas20}, which is a Bohr-type model with gravity between three rotating ultrarelativistic neutrinos as the centripetal force. This model provides a simple mode of mass generation via the relativistic increase in particle mass with increasing rotational speed.

The resulting simple equations for the rotational motion of relativistic neutrino triads, coupled with the de Broglie wavelength equation of quantum mechanics, combine, apparently for the first time, relativity and quantum mechanics \cite{Vayenas12,Research,Vayenas20} and allow for the computation of the rotational radius $r$ and of the Lorentz factor $\gamma$, thus of the rotating mass $\gamma m_o$. Thus one may compute hadron masses from neutrino masses and compare them with experiment or may compute neutrino masses from hadron masses and again compare them with experiment \cite{Vayenas12,Vayenas13,Lomonosov2,Lomonosov1,Vayenas20}. 

Thus the Rotating Lepton Model (RLM), yields the masses of composite particles (e.g. protons and neutrons) which equal K/c$^2$ where K is the kinetic energy of the rotating ultrarelativistic neutrinos which form the composite particle. Agreement with experiment regarding hadron and neutrino masses is typically 1$\%$ and 5\% respectively without the use of any adjustable parameters \cite{Vayenas12,Vayenas13,Lomonosov2,Lomonosov1,Vayenas20}.
\begin{figure}[t]
\includegraphics[width=0.90\textwidth]{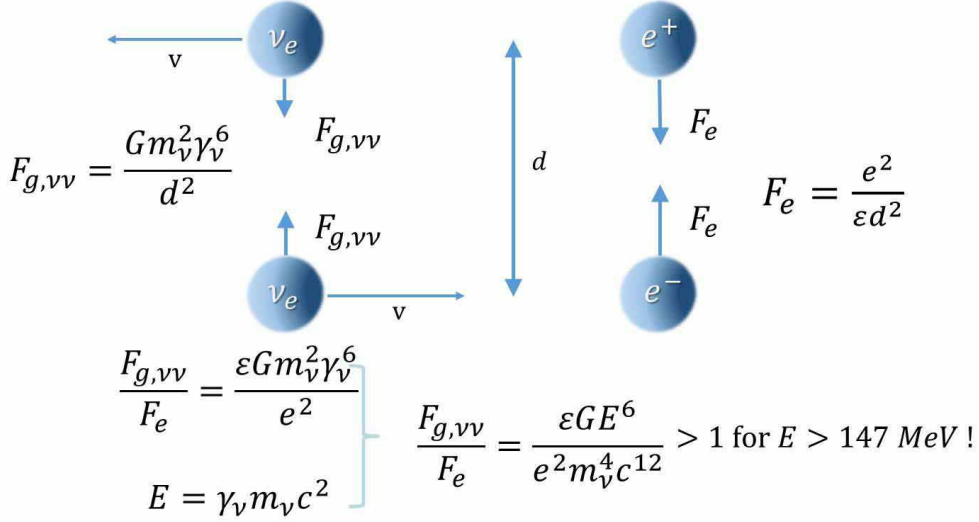}
\caption{Comparison of the gravitational force, $F_{g,\nu\nu}$, between two relativistic neutrinos with the electrostatic force, $F_e$, of a positron-electron pair at the same distance d.}
\label{fig:1}
\end{figure}

Here we first use Special Relativity coupled with the equivalence principle to compare the gravitational forces between relativistic neutrinos and between neutrino-electron pairs with Coulombic forces of electron-positron pairs at the same distance as a function of neutrino energy (Figure 2). We then present the Rotating Lepton Model (RLM) of elementary particles and we proceed to solve it for three gravitating neutrinos first via Special Relativity and then using the Schwarzschild geodesics of General Relativity. It is shown that both SR and GR lead to the same results. We also compare the computed via the RLM masses of fifteen composite particles (hadrons and bosons) with the experimental masses and we find a surprisingly good, semiquantitative, ($\pm$ 1\%) agreement. Finally we analyze why neutrinos, due to their very small rest masses, have, at fixed energy, the highest hadronization propensity among other heavier elementary particles. 

\section{Comparison of the gravitational and electrostatic forces. The Newton-Einstein gravitational equation}
Although gravitational forces between two neutrinos at rest are very weak in relation with the electrostatic attraction of a positron-electron pair at the same distance, the situation changes dramatically when the two neutrinos have a large velocity, v, with respect to a laboratory observer. This is demonstrated in Figure 1 where the gravitational force, $F_{g,\nu\nu}$, between two neutrinos with a speed v is compared with the electrostatic force, $F_e$, of a positron-electron pair at the same distance. We thus consider the ratio 
\begin{equation}
\rho_{ge}=\frac{F_{g,\nu\nu}}{F_e}
	\label{1}
\end{equation}
and we employ Coulomb's law to express $F_e$ and Newton's Universal gravitational law, using the gravitational masses, $m_g$, of the two neutrinos to express $F_{g,\nu\nu}$, i.e. 
\begin{equation}
\rho_{ge}=\frac{(Gm^2_g/d^2)}{(e^2/\epsilon d^2)}=\frac{\epsilon G m^2_g}{e^2}
	\label{2}
\end{equation}

It is worth noting that the equality
\begin{equation}
F_g=\frac{Gm^2_g}{d^2}
	\label{3}
\end{equation}
used in (\ref{2}) is equivalent with the definition of the gravitational mass, $m_g$, of a particle, which is defined as the mass value which when used in Newton's Universal gravitational law, gives the actual gravitational force $F_g$, value, \cite{Lomonosov1} i.e. 
\begin{equation}
	m^2_g=\frac{F_gd^2}{G}
	\label{4}
\end{equation}

According to the equivalence principle \cite{Roll1964} the gravitational mass, $m_g$, is equal to the inertial mass, $m_i$, and for linear motion the latter can be computed using the special relativistic definitions of force and of momentum i.e. $p=\gamma m_\nu\texttt{v}$ where $\gamma$ is the neutrino Lorentz factor $(\gamma=(1-\texttt{v}^2/c^2)^{-1/2})$, and $m_\nu$ is the neutrino rest mass, to obtain \cite{Vayenas12,Einstein1905,French68}
\begin{eqnarray}
	\nonumber F&=&\frac{dp}{dt}=\frac{d(\gamma \texttt{v}m_{\nu})}{dt}=m_\nu\left[\gamma+\texttt{v}\frac{d\gamma}{d\texttt{v}}\right]\frac{d\texttt{v}}{dt}=\\
\nonumber &=&m_{\nu}\left[\gamma+\texttt{v}\frac{d\left(1-\frac{\texttt{v}^2}{c^2}\right)^{-1/2}}{d\texttt{v}}\right]\frac{d\texttt{v}}{dt}=m_{\nu}\left[\gamma+\frac{\texttt{v}^2}{c^2}\gamma^3\right]\frac{d\texttt{v}}{dt}=\\
&=&m_{\nu}\left[\gamma+\left(1-\frac{1}{\gamma^2}\right)\gamma^3\right]\frac{d\texttt{v}}{dt}=m_{\nu}\gamma^3\frac{d\texttt{v}}{dt}
	\label{5}
	\end{eqnarray}

Since for linear motion the inertial particle mass, $m_i$, is defined from 
\begin{equation}
	m_i=F/(d\texttt{v}/dt)
	\label{6}
\end{equation}
it follows
\begin{equation}
	m_i=\gamma^3m_\nu
	\label{7}
\end{equation}
and therefore from the equivalence principle \cite{Roll1964} one obtains
\begin{equation}
	m_g=m_i=\gamma^3 m_\nu
	\label{8}
\end{equation}

Using instantaneous reference frames \cite{Vayenas12,French68} one can show that equation (\ref{8}) remains valid for arbitrary particle motion, including circular motion.

Thus using equation (\ref{8}) in equation (\ref{2}) one obtains
\begin{equation}
\rho_{ge}=\frac{eGm^2_\nu\gamma^6}{e^2}
	\label{9}
\end{equation}

Upon recalling the Einstein equation
\begin{equation}
E=\gamma m_\nu c^2
	\label{10}
\end{equation}
to express the total (kinetic plus rest) energy of the neutrino one obtains from (\ref{9})
\begin{equation}
\rho_{ge}=\frac{F_{g,\nu\nu}}{F_e}=\frac{\epsilon G}{e^2 c^{12}}\left(\frac{E^6}{m^4_\nu}\right)
	\label{11}
\end{equation}
\begin{figure}
\includegraphics[width=0.80\textwidth]{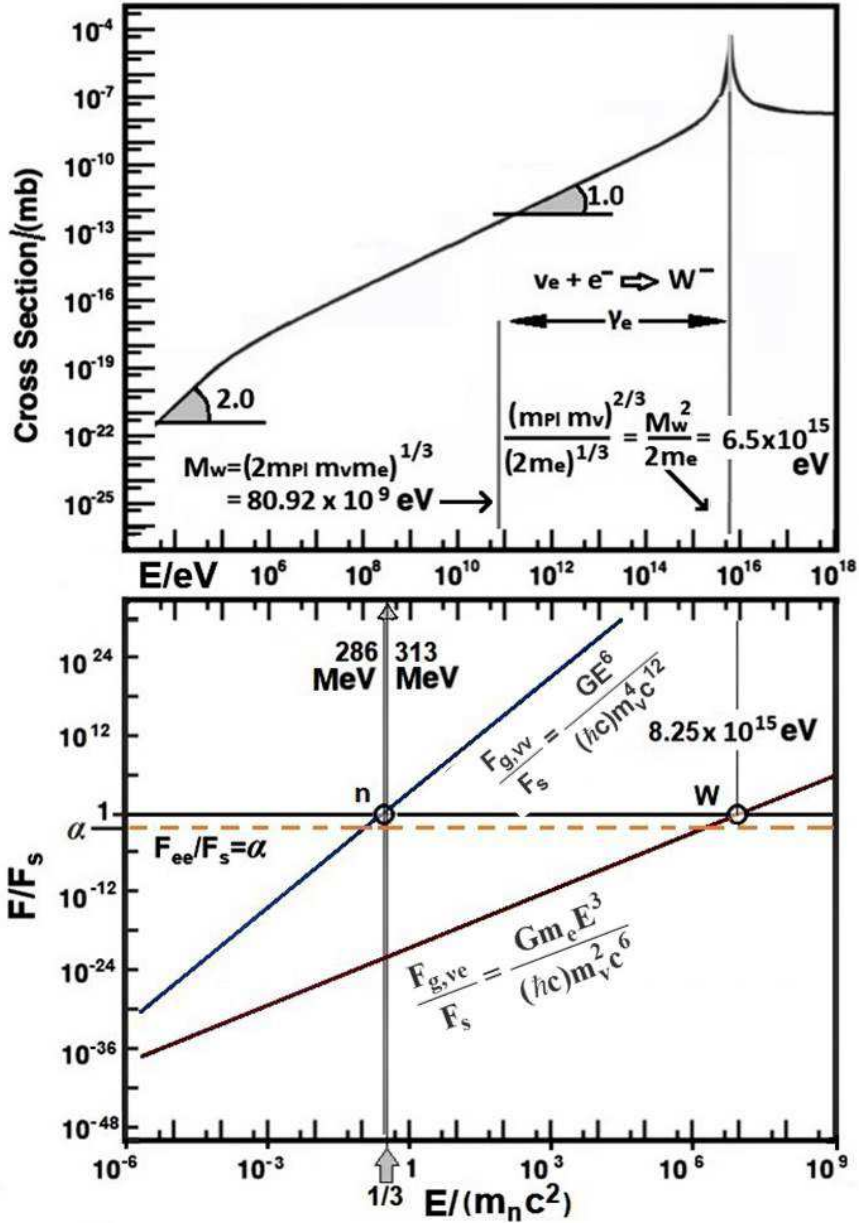}
\caption{Dependence on neutrino energy $E$, of the experimentally measured cross section for $ \bar{\nu}_e e^- \rightarrow \bar{\nu}_e e^-$ scattering \cite{Formaggio12} (top) and comparison (bottom) of the dependence on $E$ at any fixed distance, of the ratios of Coulombic force $F_{ee}$ of a positron electron pair and of the gravitational forces between two neutrinos, $F_{g,\nu\nu}$, and between an electron and a neutrino, $F_{g.\nu e}$, all divided by the strong force value $F_S=\hbar c/d^2$, computed from equations (\ref{13}), (\ref{14}) and (\ref{15}); $\alpha=e^2/\epsilon c \hbar=1/137.035$; Points n and W correspond to the energy of formation (hadronization) of neutron forming quarks with effective mass $m_q=m_n/3=313$ MeV/c$^2$) and of W bosons (from cosmic neutrinos and terrestrial $e^\pm$, the latter at rest with the observers on earth, see Appendix 1).}
\label{fig:2}
\end{figure}

By introducing $\epsilon=1.112\cdot10^{-10}$ $C^2/Nm^2$, $G=6.676\cdot 10^{-11}$ $m^3/kg s^2$, $e=1.602\cdot 10^{-19}$ $C$, $m_{\nu}=m_3 \approx 0.0437$ $eV/c^2$=$7.79\cdot 10^{-38}$ $kg$ for the heaviest neutrino \cite{Vayenas12,Mohapatra07}, one finds that, as shown in Figures (1) and (2), for $m_\nu \approx 0.0437$ eV/c$^2$ \cite{Vayenas12,Mohapatra07}, the ratio $\rho_{ge}$ exceeds unity for $E>147 $ MeV, which is of the order of the quark effective energy.

As also shown by equation (\ref{11}) the ratio $\rho_{ge}$ increases dramatically with decreasing particle mass $m_\nu$. Thus for E=147 MeV, the ratio $\rho_{ge}$ exceeds unity for $m_\nu < 0.0437$ $eV$. This is why, due to their small rest mass, neutrinos are ideal building elements of composite particles and thus, are ideal building elements of our Universe \cite{Research}. 

Figure 2 presents plots of the following forces as a function of the neutrino energy E:\\

1. The gravitational force, $F_{g,\nu\nu}$ between 2 neutrinos, where
\begin{equation}
F_{g,\nu\nu}=Gm^2_o\gamma^6/d^2=\frac{G E^6}{m^4_\nu c^{12} d^2}
	\label{12}
\end{equation}

2. The gravitational force, $F_{g,\nu e}$, between an electron/positron and a neutrino, where
\begin{equation}
F_{g,\nu e}=\frac{Gm_em_\nu\gamma^3}{d^2}=\frac{G E^3 m_e}{m^2_\nu c^6 d^2}
	\label{13}
\end{equation}

In figure 2 these forces are plotted vs the neutrino energy $E(=\gamma m_\nu c^2)$ and are compared with\\
A. The electrostatic force $F_e$ of a positron-electron pair $e^2/\epsilon d^2$\\
B. The strong force expression \cite{Griffiths08,Vayenas12}
\begin{equation}
F_{S}=\frac{\hbar c}{d^2}
	\label{14}
\end{equation}

The key features of Figure 2 are the following:\\
a. The neutrino-neutrino gravitational force, $F_{g,\nu\nu}$, reaches the strong force, $F_S$, at point n (neutron), i.e. at 
\begin{equation}
E=m_qc^2=m_n c^2/3=313\;MeV
	\label{15}
\end{equation}
where $m_q$ is the effective quark mass and $m_n$ is the neutron mass. Consequently, as shown in section 3, the point n corresponds to neutron formation, i.e. to hadronization (or baryogenesis).\\
b. The neutrino-electron (positron) gravitational force $F_{g,\nu e}$ reaches the strong force value, $F_S$, at point W, (Fig. 2 bottom).

This point W practically coincides with the Glashow resonance \cite{Formaggio12,Sarira18} (Figure 2 top) and, as shown in Appendix 1, is located at the energy
\begin{equation}
E=\frac{(m_{Pl}/m_\nu)^{2/3}}{(2m_e)^{1/3}}=6.53\cdot 10^{15} eV=6.53 PeV
	\label{16}
\end{equation}
which is the energy computed via the Rotating Lepton Model (RLM) \cite{Tully11} for the formation of a W boson from a terrestrial $e^\pm$ and a cosmic neutrino as depicted in Fig. 2 top and shown in Appendix 1.

\section{Computation via SR of the masses of hadrons generated via gravitational confinement of neutrinos}

In the case of neutrons and protons one considers three self-gravitating rotating neutrinos (Figure 3) and combines the special relativistic equation for circular motion \cite{Vayenas12,Vayenas13}
\begin{equation}
	F=\gamma m_\nu\texttt{v}^2/r
	\label{17}
\end{equation}
with equation (\ref{3}), also using equation (\ref{8}), to obtain
\begin{equation}
	F=\frac{Gm^2_\nu \gamma^6}{\sqrt{3}r^2}
		\label{18}
\end{equation}
where the $\sqrt{3}$ factor comes from the equilateral triangle geometry (Fig. 3).
\begin{figure}
\includegraphics[width=0.90\textwidth]{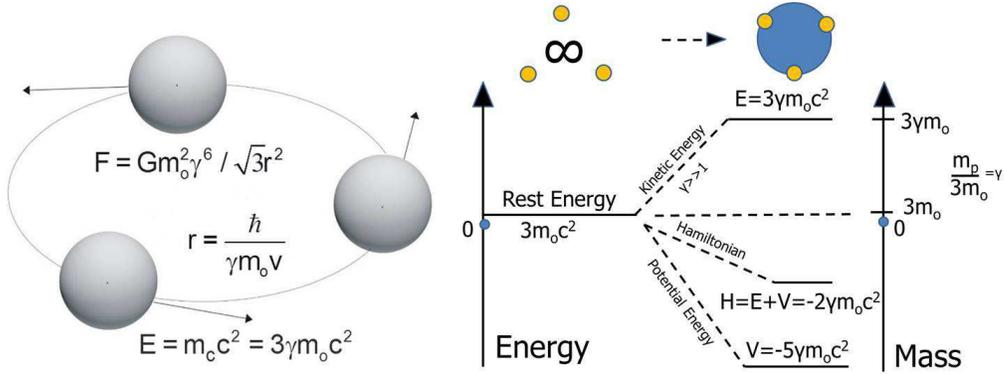}
\caption{Rotating Lepton Model (RLM) of a neutron formed by three neutrinos (left) and corresponding energy and composite mass diagram (right) showing the mechanism of gravitational particle confinement and mass generation. The composite particle (neutron) mass $3\gamma m_\nu$ is a factor of $\gamma$ larger than the rest mass $3m_\nu$ of the three constituent neutrinos. The computed via equation (22) value of $\gamma$ is 7.163$\cdot 10^9$, consequently neutrino trapping in such high $\gamma$ relativistic rotational rings is the RLM mechanism for mass generation.}
\label{fig:3}
\end{figure}

Combining (\ref{17}) and (\ref{18}) one obtains
\begin{equation}
	2\sqrt{3}\frac{r}{r_s}=\frac{\gamma^7}{\gamma^2-1}
		\label{19}
\end{equation}
with
\begin{equation}
r_s=2Gm_\nu/c^2
		\label{20}
\end{equation}
which is the Schwarzschild radius of a particle with rest mass $m_\nu$.

Equation (\ref{19}) contains two unknowns, i.e. $r$ and $\gamma$. A second equation is obtained using the de Broglie wavelength expression of quantum mechanics, as in the H Bohr model \cite{Vayenas12}, i.e.
\begin{equation}
\mathchar'26\mkern-10mu\lambda=r=\frac{\hbar}{\gamma m_\nu \texttt{v}}
		\label{21}
\end{equation}

Solution of equations (\ref{19}) and (\ref{21}), which account, respectively, both for relativity and for quantum mechanics, gives:
\begin{equation}
\gamma= 3^{1/12} (m_{Pl}/m_\nu)^{1/3}=7.163\cdot 10^{9}\quad ; \quad \gamma^3m_\nu=3^{1/4}m_{Pl}
\label{22}
\end{equation}
\begin{equation}
m_n=3\gamma m_\nu=3^{13/12}(m_{Pl}m^2_\nu)^{1/3}=939.56 MeV/c^2
\label{23}
\end{equation}
\begin{equation}
F_s/F_N=\gamma^6=3^{1/2}({m_{Pl}}/{m_\nu})^2\quad ; \quad r=\frac{3\hbar}{m_pc^2}=0.630fm
\label{24}
\end{equation}
where $m_{Pl}$ is the Planck mass $(\hbar c/G)^{1/2}=1.221\cdot 10^{19}$ GeV/c$^2$ and $F_N$ denotes the classical Newtonian $(\gamma=1)$ force value.

For $m_\nu=m_3 \approx 0.0437 eV/c^2$, where $m_3$ denotes the heaviest neutrino mass eigenstate in the Normal Hierarchy \cite{Mohapatra07}, these results are in quantitative agreement with experiment \cite{Vayenas12,Vayenas20}.

\section{Mechanism of mass generation and computed masses of selected Baryons and Bosons}
Equations (\ref{22}) and (\ref{23}) imply that the mass of the composite (neutron) state formed by the three rotating neutrinos is a factor of $\gamma=7.163\cdot 10^9$ larger than the initial three neutrino mass of $3\times(0.0437)$ eV/c$^2$. 

Figure 3 provides a direct explanation for this impressive phenomenon. Gravity maintains the three rotating particles at a highly relativistic speed $(\texttt{v}\sim c$, $\gamma=7.163\cdot 10^9)$, so that the composite particle mass $3\gamma m_\nu$ is a factor of $\gamma$ larger than that of the three neutrinos at rest, i.e. $3m_\nu$. It is worth noting, in the right part of Figure 3, the spontaneous generation of neutrino kinetic energy, which constitutes the rest energy of the neutron formed, and the corresponding spontaneous generation of negative potential energy $-5\gamma m_\nu c^2$ \cite{Vayenas12,Vayenas13,Vayenas20} which leads to a negative Hamiltonian, $\mathcal{H}$, and thus to a stable composite particle.

\begin{table}
\begin{center}
\caption{RLM computed masses and experimental masses of composite particles \cite{Vayenas20}. $m_3$, $m_2$ and $m_1$ are the masses of the neutrino mass eigenstates $\nu_3$, $\nu_2$ and $\nu_1$ respectively, in the Normal Hierarchy \cite{Mohapatra07}. $A=(6/7)\{4\left[3^{-1/2}+3^{-3/2}+2 \cdot 6^{-3/2}\right]^{-1/6}+3\left[2^{1/2}+1/4\right]^{-1/6}\}=5.848$ \cite{Aifantis21}}.
\begin{tabular}{|p{0.5in}|p{0.7in}|p{2.1in}|p{0.8in}|p{1.8in}|} \hline 
\textbf{Ref.} & \textbf{Particle} & \textbf{Formula} & \textbf{Computed mass/ (eV/c${}^{2}$)} & \textbf{Experimental mass value / (eV/c${}^{2}$)} \\ \hline \multicolumn{5}{|p{5.4in}|}{\textbf{LEPTONS mass / (MeV/c$^{2}$)}} \\ \hline 
 & e &   &  & $0.511\cdot 10^{6} $ \\ \hline 
 & $\nu_{3}$ &   &  & $0.0437$ \\ \hline 
 & $\nu_{2}$, $\nu_1$ &   &  & $0.00695,{\rm \; }1.11\cdot 10^{-3} $ \\ \hline 
\multicolumn{5}{|p{5.4in}|}{\textbf{BARYONS mass / (MeV/c$^{2}$)}} \\ \hline 
\cite{Vayenas12,Vayenas13} & p & $3^{13/12} (m_{Pl} m_3^{2} )^{1/3} $ &  & $938.272$ \\ \hline 
\cite{Vayenas12,Vayenas13} & n & $3^{13/12} (m_{p} m_{3}^{2} )^{1/3} $ & $939.565$ & $939.565$ \\ \hline 
\cite{Vayenas20,VayenasCG16,Vayenas18} & $\Lambda$ & $\left[n_{B}^{2} (2\ell _{B} +1)\right]^{1/6} m_{p} \newline {n}_{B}=1\; ;\;\ell _{B}=1$ & $1127$ & $1116$ \\ \hline 
\cite{Vayenas20,VayenasCG16,Vayenas18} & $\Delta$ & $\left[n_{B}^{2} (2\ell _{B} +1)\right]^{1/6} m_{p} \newline n_{B}=1\; ;\; \ell _{B}=2$ & $1228$ & $1232$ \\ \hline 
\cite{Vayenas20,VayenasCG16,Vayenas18} & $\Xi ^{-} ,\Xi ^{o} $  & $\left[n_{B}^{2} (2\ell _{B} +1)\right]^{1/6} m_{p} \newline n_{B} =1{\rm \; ;\; }\ell _{B} =3$ & $1300$ & $1318$ \\ \hline 
\cite{Vayenas20,VayenasCG16,Vayenas18} & $\Sigma ^{*} $  & $\left[n_{B}^{2} (2\ell _{B} +1)\right]^{1/6} m_{p} \newline n_{B} =1{\rm \; };{\rm \; }\ell _{B} =4$ & $1356$ & $1384$ \\ \hline 
\cite{Vayenas20,VayenasCG16,Vayenas18} & $\Sigma ^{-} ,\Sigma ^{0} ,\Sigma ^{+} $  & $\left[n_{B}^{2} (2\ell _{B} +1)\right]^{1/6} m_{p} \newline n_{B}=2{\rm \; };{\rm \; }\ell _{B} =0$ & $1183$ & $1192$ \\ \hline 
 &  & $\left[n_{B}^{2} (2\ell _{B} +1)\right]^{1/6} m_{p} \newline n_{B}=2{\rm \; ;\; }\ell _{B} =1$ & $1420$ &         - \\ \hline 
\cite{Vayenas20,VayenasCG16,Vayenas18} & $\Xi ^{*,-} ,\Xi ^{*,0} $  & $\left[n_{B}^{2} (2\ell _{B} +1)\right]^{1/6} m_{p} \newline n_{B}=2{\rm \; };{\rm \; }\ell _{B} =2$ & $1547$ & $1532$ \\ \hline 
\cite{Vayenas20,VayenasCG16,Vayenas18} & $\Omega $  & $\left[n_{B}^{2} (2\ell _{B} +1)\right]^{1/6} m_{p} \newline n_{B}=2{\rm \; };\ell _{B} =3$ & $1636$ & $1672$ \\ \hline 
\multicolumn{5}{|p{5.4in}|}{\textbf{PSEUDOSCALAR MESONS and MUON mass / (MeV/c${}^{2}$)}} \\ \hline 
\cite{Lomonosov1,Vayenas20} & $\mu$ & $2^{1/3} (m_{Pl} m_{2}^{2})^{1/3} $ & $105.66$ & $105.66$ \\ \hline 
\cite{Lomonosov1,Vayenas20} & $\pi$ & $(1/2)3^{13/12} (m_{Pl} m_{2}^2)^{1/3} $ & $137.82$ & 134.98 -- 139.56 \\ \hline 
\cite{Aifantis21}  & K & $A(m_{Pl} m_{2}^{2})^{1/3}$+$m_e$ & $491.0$ & $493.67$\\ \hline 
\multicolumn{5}{|p{5.2in}|}{\textbf{BOSONS mass / (GeV/c${}^{2}$)}} \\ \hline 
\cite{Vayenas16} & W$^{\pm } $ & $2^{1/3} (m_{Pl} m_{e} m_{3} )^{1/3} $ & $81.74$ & $80.42$ \\ \hline 
\cite{Fokas16} & Z & $2^{1/2} (m_{Pl} m_{e} m_{3} )^{1/3} $ & $91.72$ & $91.19$ \\ \hline 
\cite{Fokas18} & H $^{o}$  & $2\left(\frac{1-\alpha /4}{2^{1/2} +2^{-1} } \right)^{1/6} (m_{Pl} m_{e} m_{3} )^{1/3} $ & $125.7$ & $125.1$\newline $\alpha =e^{2} /\varepsilon c\hbar \approx 1/137.035$ \\ \hline 
\end{tabular}
\label{Tabl:1}
\end{center}
\end{table}

Figure 4 and Table 1 compare the experimental and computed values for 15 composite particles, i.e. nine hadrons, three mesons and three bosons without any adjustable parameters \cite{Vayenas20}. Agreement is within 2$\%$. As already discussed and summarized in this figure, the masses of baryons are of the order of $3(m_{Pl}m^2_{3})^{1/3}\approx 0.9 $ GeV, those of mesons are of the order of $3(m_{Pl}m_{2}^2)^{1/3}\approx 0.3$ GeV, while those of bosons, which are rotational $e^\pm-\nu_e$ structures \cite{Vayenas16,Fokas16,Fokas18}, are of the order of $(2m_{Pl}m_em_{3})^{1/3}\approx 80$ GeV.

\begin{figure}
\begin{center}
\includegraphics[width=0.80\textwidth]{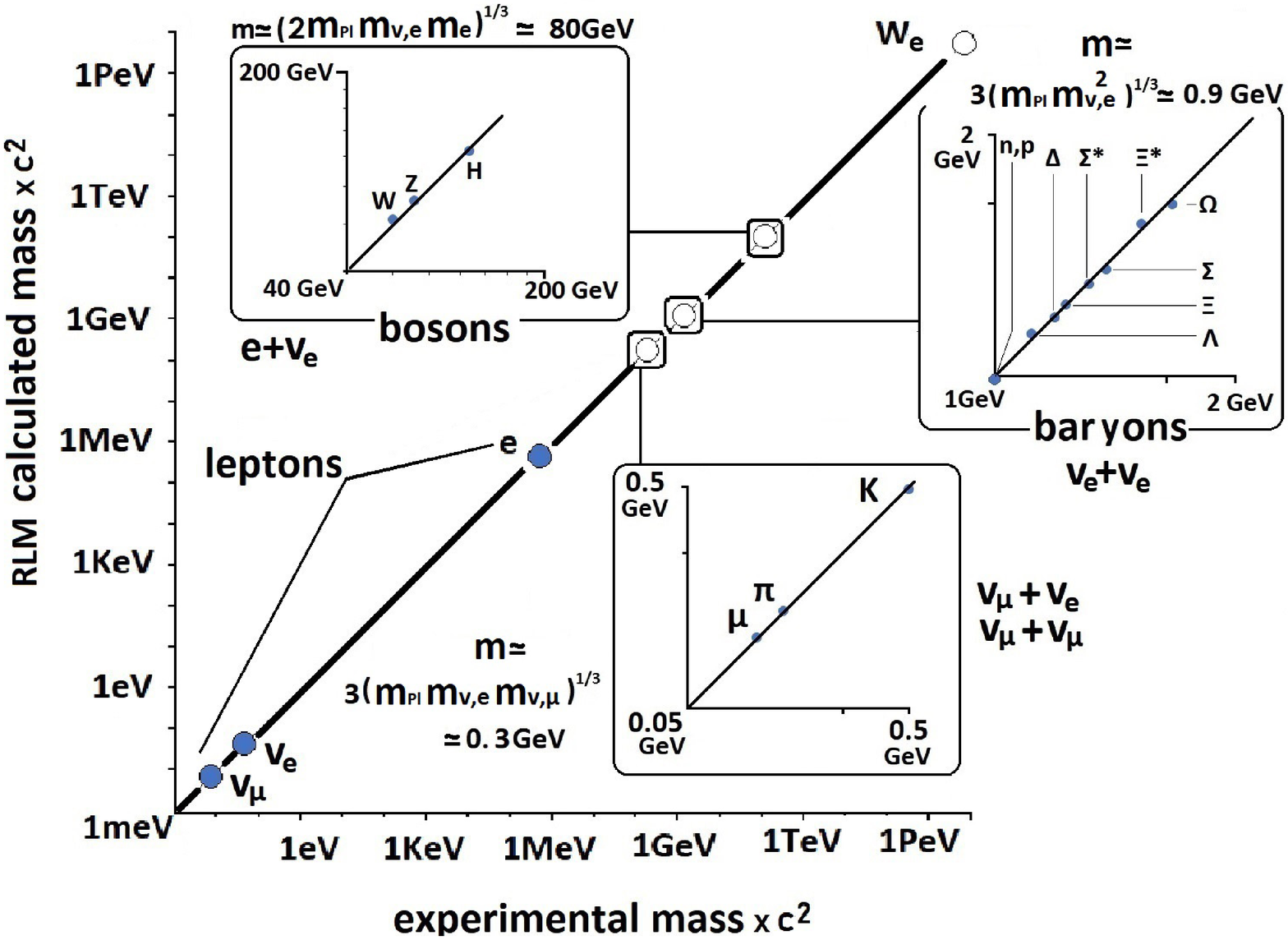}
\caption{Comparison of the RLM computed masses of composite particles with the experimental values also shown in Table 1. Agreement is better than 2\% without any adjustable parameters. The three approximate mass expressions shown in the Figure provide the order of magnitude of hadron and boson masses \cite{Vayenas20}.}
\end{center}
\label{fig:4}
\end{figure}

\section{GR treatment}
It is interesting to examine if the key results of the SR treatment, e.g. equations (\ref{23}) and (\ref{24}), can be obtained using the theory of general relativity (GR) where the masses are considered fixed and not velocity dependent \cite{Vayenas13,Wald84}. 

In order to apply the Schwarzschild geodesics equation of GR \cite{Wald84} to the rotating neutrino problem one must adjust the physical model of Fig. 3 to the standard geometry of the Schwarzschild metric which involves a light test particle of mass $m^*$ rotating around a central mass $M$ \cite{Vayenas13,Wald84}. This can be done via the model shown in Figure 5. First we note that in the three-rotating particle model the force exerted to each particle is given by $F_N=Gm^2_\nu/(\sqrt{3}r^2)$, therefore the Newtonian potential energy due to the other two particles is given by 
\begin{equation} 
\label{25} 
V_N=-Gm^2_\nu/(\sqrt{3}r).
\end{equation} 
\begin{figure}[ht]
\begin{center}
\includegraphics[width=0.40\textwidth]{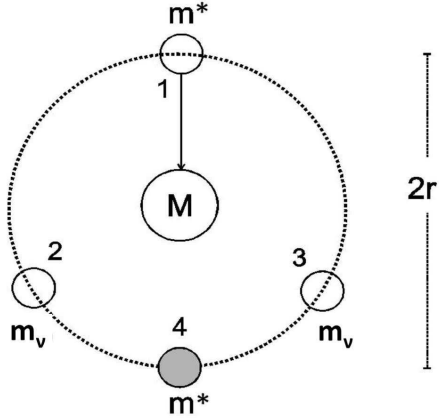}
\caption{GR Schwarzschild geodesics model for the three-neutrino state; $m^*=2^{1/2}3^{-1/4}m_\nu$  \cite{Vayenas13,Lomonosov2,Vayenas15}.} 
\label{fig:5}
\end{center}
\end{figure}

Since the same cyclic motion of particle 1 due to particles 2 and 3 can be obtained by substituting particle 1 with a new particle of mass $m^*$ and by also substituting particles 2 and 3 by a particle 4 of mass $m^*$ in the antidiametric position of 1, it follows that $V_N$ must also equal the potential energy of 1 due to particle 4, i.e. $V_N=-Gm^{*2}/2r$. From this and equation (\ref{25}) it follows $m^*=2^{1/2}3^{-1/4}m_\nu$. We then consider the one-dimensional Schwarzschild  effective potential, $V_s(r)$ with $M>>m^*$
\begin{equation} 
\label{26} 
V_s(r)=-\frac{GMm^*}{r}+\frac{2^{-1/2}3^{1/4}L^2}{2m_\nu r^2}-\frac{2^{-1/2}3^{1/4}GML^2}{c^2m_\nu r^3},
\end{equation} 
where $L$ is the angular momentum. Setting $L=\hbar$ one obtains
\begin{equation} 
\label{27} 
\frac{V_s(r)}{(m^*c^2/2)}=-\frac{r_s}{r}+\frac{a^2}{r^2}-\frac{a^2r_s}{r^3},
\end{equation}
where $r_s(=2GM/c^2)$ is the Schwarzschild radius of the central mass $M$ and $a(=\hbar/m^*c)$ is the Compton wavelength of the rotating mass $m^*$. 

The effective potential is not the actual potential experienced by the rotating particle. It is the potential of an identical particle in ordinary one-dimensional nonrelativistic mechanics which causes the same one-dimensional motion as the radial motion of the actual rotating particles \cite{Wald84,Vayenas15}. The derivative $F=dV_s/dr$, termed effective force, is thus the force acting on this identical particle in ordinary one-dimensional nonrelativistic motion. 
 
Two circular orbits are obtained when the effective force $F_s=dV_s(r)/dr$ is zero, which upon differentiation of (\ref{27}) gives 
\begin{equation} 
\label{28} 
r^\pm=\frac{a^2}{r_s}\left[1\pm \sqrt{1-\frac{3r^2_s}{2a^2}}\right].
\end{equation}

The larger one of this roots (a minimum in $V_s$) is very large $(\sim 10^{24}m)$ and irrelevant in the present model. The smaller one (a maximum in $V_s$) is unstable (Fig. 6). 

When the rotational radius $r$ is smaller than $r^-$, which is close to the Schwarzschild radius, $r_s$, then the effective force suggests that the formation of a black hole with an event horizon at $r_s$ is energetically favored (Fig. 6).

\begin{figure}[ht]
\begin{center}
\includegraphics[width=0.70\textwidth] {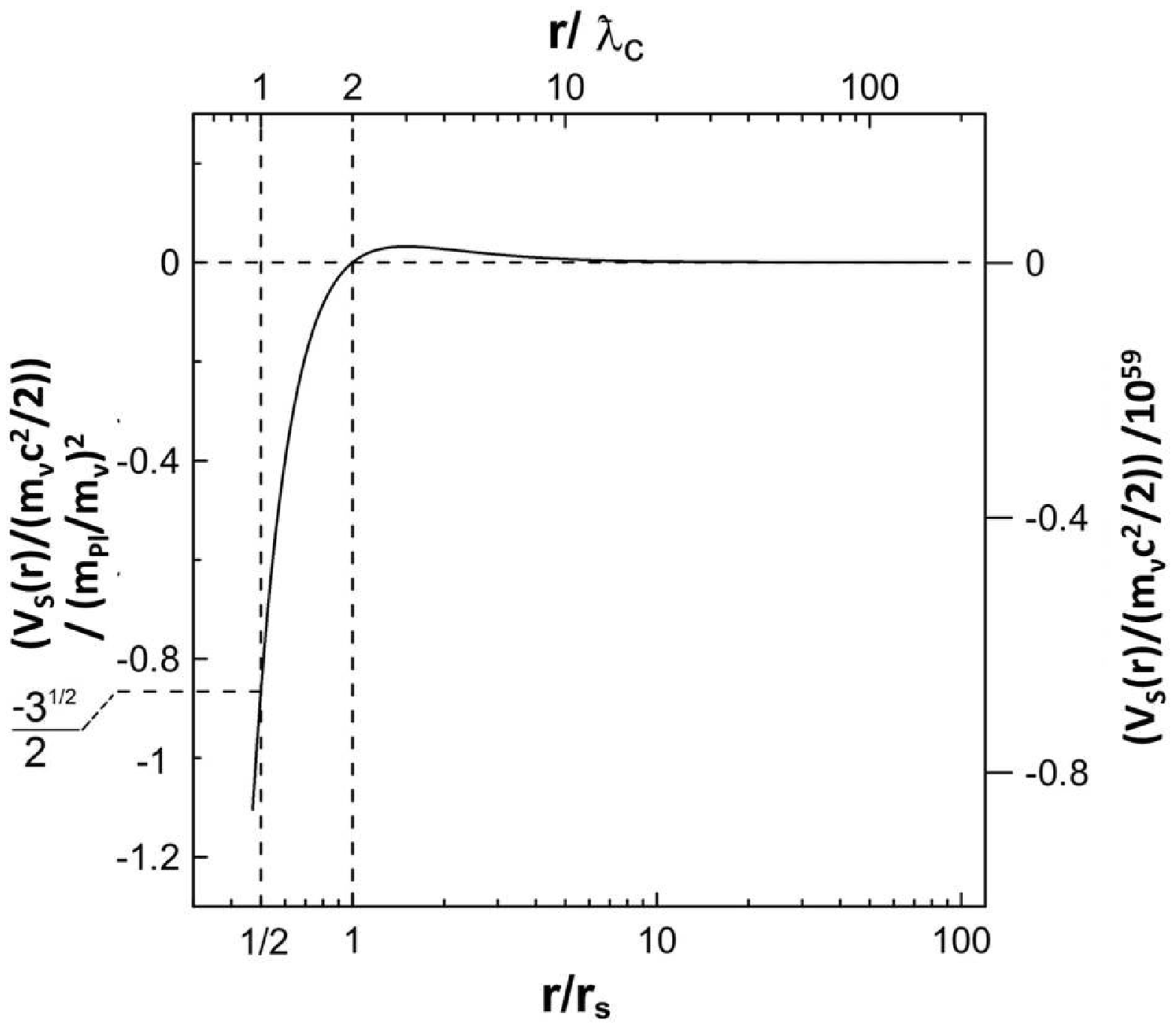}
\caption{Plot of equation (\ref{27}); $r_s=2GM/c^2$, $\mathchar'26\mkern-10mu\lambda_c=\hbar/m^*c$. }
\label{fig:6}
\end{center}
\end{figure}

Differentiation of eq. (\ref{27}) gives
\begin{equation} 
\label{29} 
F_s(r)=dV_s(r)/dr=(m^*c^2/2)\left[\frac{r_s}{r^2}-\frac{2a^2}{r^3}+\frac{3a^2r_s}{r^4}\right].
\end{equation}

The first term in (\ref{29}) corresponds to normal gravitational attraction according to Newton's gravitational law. Indeed, denoting this force by $F_N$ and accounting for $r_s=2GM/c^2$ one obtains
\begin{equation} 
\label{30} 
F_N(r)=\frac{m^*c^2}{2}\cdot \frac{2GM}{c^2r^2}=\frac{GMm^*}{r^2}.
\end{equation}

The ratio $F_s(r)/F_N(r)$ of the effective force $F_s(r)$ to the Newtonian force $F_N$ is thus given by:
\begin{equation} 
\label{31} 
F_s(r)/F_N(r)=1+\frac{a^2}{r^2_s}\left[3\left(\frac{r_s}{r}\right)^2-2\left(\frac{r_s}{r}\right)\right].
\end{equation}

Using the definition of $a$, $r_s$ and $m_{Pl}(=\hbar c/G)^{1/2}$ and recalling $m^*=2^{1/2}3^{-1/4}m_o$ we obtain
\begin{equation} 
\label{32} 
F_s(r)/F_N(r)=1+\frac{m^4_{Pl}}{4M^2m^{*2}}\left[3\left(\frac{r_s}{r}\right)^2-2\left(\frac{r_s}{r}\right)\right]=1+\frac{\sqrt{3}}{8}\cdot \frac{m^4_{Pl}}{M^2m^2_\nu}\left[3\left(\frac{r_s}{r}\right)^2-2\left(\frac{r_s}{r}\right)\right].
\end{equation}

The minimum $r$ value for which $F_N(r)$ can serve as a Newtonian centripetal force, and thus for which the ratio $F(r)/F_N(r)$ is defined, is the value $r_s/2$. This follows from 
\begin{equation} 
\label{33} 
\frac{GMm^*}{r^2}=\frac{m^*\texttt{v}^2}{r}\quad thus\quad r=\frac{GM}{\texttt{v}^2}.
\end{equation}

Accounting for $\texttt{v}< c$ it follows that indeed $r>r_s/2$. This is the minimum allowed $r$ value which leads (Fig. 6) to the lowest possible value of $V_s(r)$ and which according to eq. (\ref{27}) and accounting for $\alpha >>r_s$ is given by 
\begin{equation} 
\label{34} 
V_s(r)_{min}=(m^*c^2/2)\left[-2-\frac{4a^2}{r^2_s}\right]\approx -(m^*c^2/2)\frac{4a^2}{r^2_s}=-m^*c^2\frac{m^4_{Pl}}{2M^2m^{*2}}=-m^*c^2\frac{\sqrt{3}}{4}\cdot \frac{m^4_{Pl}}{M^2m^2_\nu}
\end{equation}

When the limit $r=r_s/2$ coincides with the Compton wavelength limit, $\mathchar'26\mkern-10mu\lambda_c(=\hbar/M c)$, imposed by the Heisenberg uncertainty principle (HUP) \cite{Das09}, i.e. when 
\begin{equation} 
\label{35} 
r_s/2=\frac{GM}{c^2}=\frac{\hbar}{Mc}=\mathchar'26\mkern-10mu\lambda_c,
\end{equation}
it follows 
\begin{equation} 
\label{36} 
M=\left(\frac{\hbar c}{G}\right)^{1/2}=m_{Pl}
\end{equation}

This result is consistent with that extracted from the $SR$ treatment (eq. \ref{22}) which shows that the mass keeping, via its gravitational attraction, each rotating particle to its orbit is the Planck mass $m_{Pl}$. 

Substituting in (\ref{32}) with $r=r_s/2$ and accounting for (\ref{36}) one obtains
\begin{equation} 
\label{37} 
F_s(r)/F_N=3^{1/2}\left(\frac{m_{Pl}}{m_\nu}\right)^2
\end{equation}
which, surprisingly, is the same result as that obtained via the special relativistic treatment in equation (\ref{24}).

It thus appears that $r_s/2$ provides a satisfactory generalized event horizon (GEH) \cite{Das09}, at least when used in conjunction with the Heisenberg uncertainty principle (HUP). At $r=r_s/2=\mathchar'26\mkern-10mu\lambda_c$ both the Schwarzschild geodesics and the Heisenberg uncertainty principle are satisfied for $M=m_{Pl}$. Therefore, as in the case of the SR treatment, (eqs. (19), (20) and (21)) hadronization of neutrinos accounts simultaneously for both Relativity and Quantum Mechanics.

\section{Conclusions}
Due to the very small values of neutrino rest masses, the gravitational forces between ultrarelativistic neutrinos are, at fixed energy, stronger than those between any other particles, and play a very significant role in hadronization. At a neutrino total energy $E(=\gamma m_\nu c^2)$ equal to $(m_n/3)c^2=313$ MeV, where $m_n$ is the neutron mass (939.585 MeV/c$^2$), the inertial, thus also gravitational, mass of a neutrino $(=\gamma^3 m_\nu)$ reaches the value of Planck mass $(=\hbar c/G)^{1/2}=1.221\cdot 10^{28}$ eV/c$^2$ and the gravitational force between two such neutrinos $(=Gm^2_{Pl}/d^2=\hbar c/d^2)$ reaches the value of the Strong Force, i.e. $\hbar c/d^2$, with a concomitant formation of hadrons.

Thus, it follows from equations (\ref{8}) and (\ref{10}) and the condition $\gamma^3m_\nu=m_{Pl}$, that the minimum energy for hadronization, $E_h$, of a particle with rest mass $m_o$ is given by the expression
\begin{equation}
E_h=(m_{Pl}m^2_o)^{1/3}c^2
\label{38}
\end{equation}

Therefore, the lighter an elementary particle is, the smaller the energy demand for its gravitational confinement. Since neutrinos have the smallest rest mass of all known particles, it follows that they also have the highest hadronization propensity due to the ease with which their inertial and thus gravitational mass reaches the Planck mass value, $(\hbar c/G)^{1/2}$, and thus the gravitational force between them reaches the value of the Strong Force $(\hbar c/d^2)$. As equation (\ref{38}) implies, a particle with the rest mass, $m_3$ of a heavy neutrino (0.0437 eV/c$^2$) requires for its hadronization an energy $E_h=313$ MeV, while a particle with the mass of an electron (0.511 MeV/c$^2$) requires for its hadronization an energy $E_h$ of 16.1 TeV, i.e. a factor of 5.6$\cdot 10^4$ larger than that required for a neutrino. This explains why the particles with the smallest rest mass, i.e. the neutrinos, are ideal elements for hadronization, i.e. for the generation of mass $\gamma m_o$, via their high $\gamma$ circular motion, and thus for the synthesis of hadrons and other composite particles.

In summary, it appears that the ability of the RLM in computing the masses of composite particles may be attributed to the fact that in a simple manner it combines Relativity with Quantum Mechanics \cite{Research}. Such a combination has long been expected to lead to useful results \cite{Research,Weinberg,Penrose,Hawking}.

\newpage
 \renewcommand{\theequation}{1-\arabic{equation}}
  \setcounter{equation}{0}  
  \renewcommand{\thefigure}{1-\arabic{figure}}
  \setcounter{figure}{0}
	\renewcommand{\thetable}{1-\arabic{table}}
  \setcounter{table}{0}
  \section*{APPENDIX 1: W$^\pm$ boson mass and the Glashow resonance}
It has been shown recently \cite{Vayenas20,Vayenas16} via the Rotating Lepton Model (RLM) that W$^\pm$ bosons are rotating relativistic $e^\pm-\nu_e$ pairs \cite{Vayenas20}. Here we derive both expressions for the W mass, $M_W$, given in Figure 2, i.e.
\begin{equation}
\label{1-1}
M_W=(2m_{Pl}m_\nu m_e)^{1/3}=80.92\cdot 10^9\;eV
\end{equation}
which is in excellent agreement with the experimental W mass of 80.42$\cdot 10^9$ eV, as well as the expression 
\begin{equation}
\label{1-2}
m'_W=\frac{(m_{Pl}m_\nu)^{2/3}}{(2m_e)^{1/3}}=6.5\cdot 10^{15}\;eV=6.5\;PeV
\end{equation}
Equation (\ref{1-1}) is derived by considering that the observer is at rest with the center of rotation of the $e^\pm-\nu_e$ pair, which is always the case for terrestrial systems (unless their center of mass is moving towards the observer with a speed near c). Thus we have
\begin{equation}
\label{1-3}
\gamma_\nu m_\nu \texttt{v}^2/r=\gamma_e m_e \texttt{v}^2/r=Gm_e m_\nu\gamma^3_\nu\gamma^3_e/4r^2
\end{equation}
therefore
\begin{equation}
\label{1-4}
4\gamma_e m_e \texttt{v} r\texttt{v}=4\gamma_\nu m_\nu \texttt{v} r\texttt{v}=Gm_em_\nu\gamma^3_e\gamma^3_\nu
\end{equation}

Recalling the de Broglie wavelength equation, i.e.
\begin{equation}
\label{1-5}
\gamma_e m_e \texttt{v}_e r=\gamma_\nu m_\nu \texttt{v}_\nu r=\hbar
\end{equation}
and accounting for $\texttt{v}_1\approx \texttt{v}_2 \approx c$, equation (\ref{1-3}) gives
\begin{equation}
\label{1-6}
4\hbar c=Gm_e m_\nu \gamma^3_e \gamma^3_\nu
\end{equation}

Defining 
\begin{equation}
\label{1-7}
(m_em_\nu)^{1/2}=m_{e\nu} \quad ; \quad (\gamma_e \gamma_\nu)^{1/2}=\gamma_{e\nu}
\end{equation}
equation (\ref{1-6}) gives
\begin{equation}
\label{1-8}
4\hbar c=G m^2_{e\nu}\gamma^6_{e\nu}\quad ; \quad \gamma_{e\nu}=\left(\frac{4\hbar c}{Gm^2_{e\nu}}\right)^{1/6}
\end{equation}

Recalling the definition of the Planck mass $m_{Pl}=(\hbar c/G)^{1/2}$, equation (\ref{1-8}) becomes
\begin{equation}
\label{1-9}
\gamma_{e\nu}=2^{1/3}m^{1/3}_{Pl}/m^{1/3}_{e\nu}
\end{equation}
therefore the mass of each of the two isoenergetic fragments detected is: 
\begin{equation}
\label{1-10}
M_W/2=\gamma_{e\nu}m_{e\nu}=2m^{1/3}_{Pl}m^{2/3}_{e\nu}
\end{equation}
and using the first equation (\ref{1-7})
\begin{equation}
\label{1-11}
M_W=(2m_{Pl}m_em_{\nu})^{1/3}=80.92\; GeV/c^2 
\end{equation}
in astonishing agreement with the experimental value of 80.42 GeV/c$^2$.

Equation (\ref{1-2}) is derived by considering that the observer is at rest with the terrestrial electron, thus $\gamma'_e=1$. In this case we have 
\begin{equation}
\label{1-12}
\gamma'_\nu m_\nu \texttt{v}^2/2r=\frac{G m_e m_\nu \gamma'{^3}_\nu}{4r^2}
\end{equation}

Therefore in combination with
\begin{equation}
\label{1-13}
\gamma'_\nu m_\nu \texttt{v}(4r)=\hbar
\end{equation}
it follows
\begin{equation}
\label{1-14}
\hbar c=2G m_e m_\nu\gamma'_\nu{^3}
\end{equation}
therefore
\begin{equation}
\label{1-15}
\gamma_\nu'{^3}=\left(\frac{\hbar c}{2G m_e m_\nu}\right)^{1/3}=\frac{m^{2/3}_{Pl}}{2m^{1/3}_em^{1/3}_{\nu}}
\end{equation}
and thus
\begin{equation}
\label{1-16}
m'_W=\gamma'_\nu m_\nu=\frac{m^{2/3}_{Pl}}{2m^{1/3}_em^{1/3}_{\nu}}m_\nu=\frac{(m_{Pl}m_\nu)^{2/3}}{(2m_e^{1/3})}=6.53\cdot 10^{15}\;eV/c^2=6.53\;PeV/c^2
\end{equation}
which practically coincides with the experimentally observed value of 6.3 PeV for the Glashow transition \cite{Sarira18}. Therefore one may conclude that the Glashow resonance is due to the reaction
\begin{equation}
\label{1-17}
e^\pm(terrestrial)+\nu_e(cosmic\;ray)\rightarrow W^\pm
\end{equation}
occuring between cosmic ray neutrinos and terrestrial electrons or positrons. 
\end{document}